\documentclass[aps,tightenlines,nofootinbib,superscriptaddress,amsfonts,amsmath,preprintnumbers]{revtex4}

\usepackage{color}
\newcommand{\ba}{\begin{eqnarray}}
\newcommand{\ea}{\end{eqnarray}}
\newcommand{\bea}{\begin{eqnarray}}
\newcommand{\eea}{\end{eqnarray}}
\newcommand{\nn}{\nonumber}
\newcommand{\be}{\begin{equation}}
\newcommand{\ee}{\end{equation}}
\newcommand{\beq}{\begin{equation}}
\newcommand{\eeq}{\end{equation}}

\newcommand{\mrm}[1]{\mathrm{#1}}
\newcommand{\cK}{{{\cal K}}}

\newcommand{\emu}{{\rm e}^\mu}

\def\lb{\left(}
\def\rb{\right)}

\usepackage{graphicx}
\usepackage{amsmath,amssymb,bm,float,mathrsfs}
\usepackage{dcolumn}
\usepackage{color}
\usepackage{hyperref}

\begin{document}

\title{Y-junction intercommutations of current carrying strings}

\author{D.~A.~Steer} \email{steer@apc.univ-paris7.fr}
\affiliation{AstroParticule \& Cosmologie,
UMR 7164-CNRS, Universit\'e Denis Diderot-Paris 7,
CEA, Observatoire de Paris,
10 rue Alice Domon et L\'eonie
Duquet, F-75205 Paris Cedex 13, France}
%\affiliation{${\mathcal{G}}{\mathbb{R}}\varepsilon{\mathbb{C}}{\mathcal{O}}$,
%Institut d'Astrophysique de Paris,
%CNRS, UMR 7095, et Sorbonne Universit\'es,
%UPMC Univ Paris 6,
%98bis boulevard Arago, F-75014 Paris, France}

\author{Marc Lilley} \email{lilley@iap.fr}
\affiliation{${\mathcal{G}}{\mathbb{R}}\varepsilon{\mathbb{C}}{\mathcal{O}}$,
Institut d'Astrophysique de Paris,
CNRS, UMR 7095, et Sorbonne Universit\'es,
UPMC Univ Paris 6,
98bis boulevard Arago, F-75014 Paris, France}

\author{Daisuke Yamauchi}\email{yamauchi@jindai.jp}
\affiliation{Faculty of Engineering, Kanagawa University, Kanagawa, 221-8686, Japan
}

\author{Takashi Hiramatsu}\email{hiramatz@yukawa.kyoto-u.ac.jp}
\affiliation{Department of Physics, Rikkyo University, Tokyo 171-8501, Japan
}

\begin{abstract}

Under certain conditions the collision and intercommutation of two cosmic strings can result in the formation of 
a third string, with the three strings then remaining connected at Y-junctions.  The kinematics and dynamics of collisions of this type have been the subject of analytical and numerical analyses in the special case in which the strings are Nambu-Goto.  Cosmic strings, however, may well carry currents, in which case their dynamics is not given by the Nambu-Goto action. Our aim is to extend the kinematic analysis to more general kinds of string model. We focus in particular on the collision of strings described by conservative elastic string models, characteristic of current carrying strings, and which are expected to form in a cosmological context.  As opposed to Nambu-Goto strings collisions, we show that in this case the collision cannot lead to the formation of a third elastic string: if dynamically such a string forms then the joining string must be described by a more general equation of state. This process will be studied numerically in a forthcoming publication.

\end{abstract}

%\date{\today}

%\pacs{04.50.-h, 11.10.-z}

\preprint{RUP-17-20}
\maketitle

%\tableofcontents

\section{Introduction}
\label{sec:intro} 

Strings with Y-shaped junctions occur in many models, and have been the subject of numerous studies, from QCD \cite{Artru:1974zn}, to cosmic strings \cite{Bettencourt:1996qe,Copeland:2006eh,Copeland:2006if,Salmi:2007ah,Bevis:2008hg,Bevis:2009az,Hiramatsu:2013yxa,Hiramatsu:2013tga}, and cosmic superstrings \cite{Copeland:2007nv,Rajantie:2007hp,Urrestilla:2007yw,Davis:2008kg}.   In a cosmological context, the presence of Y-junctions on a string network can 
give rise to many interesting effects: for instance, the lensing pattern of a distant light source background sources can pick up a distinctive triplified aspect \cite{Shlaer:2005ry,Brandenberger:2007ae}; when kinks (discontinuities on the tangent vector of a string) travel through a junction they multiply (in number) thus potentially sourcing more gravitational waves \cite{Binetruy:2009vt,Binetruy:2010cc}; and also networks of cosmic superstrings lead to novel effects on the CMB temperature and polarization spectra \cite{Pourtsidou:2010gu,Avgoustidis:2011ax}.

Detailed analytic studies \cite{Copeland:2006eh,Copeland:2006if} of the collisions of strings and the subsequent formation of Y-junctions have, so far, focused on Nambu-Goto (NG) strings, which are idealised strings that are infinitely thin and carry no internal structure\footnote{See also \cite{Copeland:2007nv} for an extension to strings governed by the Dirac-Born-Infeld action.}.  In this paper we consider the more general case of elastic string models, which provide a macroscopic description of different kinds of current-carrying and superconducting strings such as those originally proposed by Witten in \cite{Witten:1984eb}.  Crucially, in a cosmological context, current carrying strings are expected to be formed in many supersymmetric models of inflation \cite{Jeannerot:1997is}. 
Thus while it seems  probable that cosmic strings carry currents, the collision of such strings has not been fully investigated.  That is the aim of this paper.

There are important differences between NG and current carrying strings, the latter of which carry internal degrees of freedom. From a technical point of view, in particular, the world-sheet gauge choices often made to study NG strings (the conformal and temporal gauges) do not apply to general elastic string models.  Furthermore, the equations of motion of these strings are generally not integrable (the exception being the chiral case, in which the worldsheet current is null \cite{Carter:1999hx}).  The above-mentioned gauge choices were made in previous works on NG strings \cite{Copeland:2006eh,Copeland:2006if}, thus rendering them inapplicable to the study of elastic string collisions. For this reason, here we develop a fully covariant formalism --- with respect to both the string worldsheet and the background spacetime \cite{Carter:1989dp,Carter:1989xk,Carter:1996mx} --- to study the collision of current-carrying strings with Y-junctions.  This formalism applies both to NG strings where it reduces to that of \cite{Copeland:2006eh,Copeland:2006if}, and to different elastic-string models as discussed in Sec.~\ref{sec5}.

The simplest field theory model in which Y-junctions form is the Abelian-Higgs model in the type I regime where higher winding number (labeled by the integer $n$) strings are stable. It has been shown that when two $n=1$ strings collide, under certain conditions --- which depend on the relative velocity and angle of the two colliding strings \cite{Copeland:2006eh,Copeland:2006if,Salmi:2007ah} --- a third $n=2$ string is formed with two corresponding Y-junctions joining 
it to the original 2 strings.  Elastic current-carrying strings are obtained by extending this model to a
$U(1)\times U(1)$ theory, as first discussed by Witten~\cite{Witten:1984eb}. Here the first $U(1)$ forms the strings while, if the coupling between the two $U(1)$ sectors is chosen appropriately, the second $U(1)$ can condensate on the strings
and generate a current.  This model has been studied in depth, both from a field theory (microscopic) \cite{Peter:1992dw} and effective action (macroscopic) point of view \cite{Copeland:1987th,Carter:1994hn,Hartmann:2008yr}. It is this second, effective action approach, that we follow in this paper.  The questions we are interested in are: what happens to the currents when the two strings collide? Can junctions form?  If so, what are the properties of the joining string? Initial numerical studies of collisions of this kind were presented in \cite{Laguna:1990it}; further studies will be presented in a companion paper \cite{Hiramatsu}.

This paper is set up as follows. In section \ref{sec1} we briefly review some of the main properties of elastic string models. In section \ref{sec2} we set up the necessary formalism to describe junctions between elastic strings, and derive the corresponding junction conditions. In section \ref{sec3} we apply these junction conditions to V-junctions at which two strings join; and in section \ref{sec4} we apply them to Y-junctions. String collisions are studied in section \ref{sec5}. Our conclusions are presented in section \ref{sec6}.

\section{Preliminaries: Elastic strings}
\label{sec1}

Elastic strings contain internal degrees of freedom, and (for a string with no junctions) the models we consider are governed by an effective action of the form \cite{Carter:1989xk,Carter:1989dp} 
\be
S = \int  d^2  \sigma \sqrt{-{\rm det}{(\gamma_{ab}})} \; {\cal L}(w)\, ,
\label{action}
\ee
where  $\sigma^{a} = (\tau,\sigma)$ are the internal coordinates of the string worldsheet. In terms of the string position $x^\mu(\sigma,\tau)$, the induced metric $\gamma_{ab}$ is given by
\be
\gamma_{ab} = \eta_{\mu \nu} x^\mu_{,a} x^\mu_{,b}\, ,
\nonumber
\ee
 where a comma denotes partial differentiation, and $\eta_{\mu \nu}$ is the Minkowski metric (with mostly positive signature). For NG strings, the world-sheet Lagrangian $ {\cal L}$ is constant ---  there are no internal world-sheet degrees of freedom --- and hence the string is locally Lorentz invariant. Thus the tension $T$ and energy per unit length $U$ of NG strings are equal and constant.

 For elastic strings \cite{Carter:1989xk,Carter:1989dp}, first introduced in the context of cosmic string theory for studying the mechanical effects of the
currents that occur in ``superconducting'' strings \cite{Witten:1984eb,Carter:1999hx}, the Lagrangian is a function of the variable $w$ (often referred to as the state parameter) defined by 
 \be
 w \equiv \kappa_0 \gamma^{ab} \varphi_{,a} \varphi_{,b}
 \label{wdef}
 \ee
where $\kappa_0$ is a freely adjustable positive dimensionless normalisation constant. The scalar field $\varphi(\sigma^i)$ can be viewed as a stream function associated with the conserved current\footnote{For simplicity we assume that the field is not charged under electromagnetism.}, which itself arises from the invariance of ${\cal L}(w)$ under $\varphi \rightarrow \varphi+$constant, see below.
In many physical models, $\varphi$ can be associated with a (dimensionless) phase: for instance, in a field theory with $U(1)_{\rm local}\times U(1)_{\rm global}$ symmetry, $\varphi$ can be identified with the phase of the $U(1)_{\rm global}$ field which condenses in the core of the string (which is itself formed by the breaking of the $U(1)_{\rm local}$ symmetry group).
The precise form of ${\cal L}(w)$ depends on the underlying field theory, and has been the subject of numerous studies, see for instance \cite{Carter:1994hn,Hartmann:2008yr}.  In terms of the mass  $m$ associated with the symmetry breaking scale and the mass $M$ of the current carriers,
{\it electric strings} (with time-like $w<0$)  in the $U(1)_{\rm local}\times U(1)_{\rm global}$ can be described by the Lagrangian  \cite{Hartmann:2008yr}
\be
{\cal L}(w) = -m^2 - \frac{{M}^2}{2} \ln\left(1+\frac{w}{M^2}\right)~.
\label{Le}
\ee
The divergence at $w=-M^2$ corresponds to the threshold for current carrier particle creation.
For {\it magnetic} strings for which $w$ is space-like, $w>0$, a suitable Lagrangian is  \cite{Carter:1994hn}\be
{\cal L}(w) = -m^2 - \frac{w}{2}\left(1 - \frac{w}{M^2}\right).
%{\cal L}(w) = -m^2 - \frac{w}{2}\left(1 - \frac{w}{M^2}\right)^{{-1}}.
\label{Lm}
\ee
Lightlike null currents $w=0$ 
must be treated separately, and in this case the equations of motion are integrable \cite{Carter:1999hx,Davis:2000cx,BlancoPillado:2000ep}.   
For most of this analysis we shall leave ${\cal L}(w)$ arbitrary, but we exclude chiral strings $w=0$ which will be considered elsewhere.

The  equations of motion obtained from (\ref{action}) by varying with respect to $\varphi$ and $x^\mu$ take the form
\ba
\nabla_a c^a &=& 0 \, , 
\nn
\\
\nabla_a(T^{ab} x^\mu_{,b}) &=& 0  \, , 
\nn
\ea
where the conserved current $c^a$, and the stress energy tensor $T^{ab}$, are given by
\ba
c^a  &\equiv &  \frac{\sqrt{\kappa_0}}{{\cal K}(w)} \gamma^{ab} \varphi_{,b} \, , 
\label{cdef}
\\
T^{ab} &\equiv & {\cal L}(w) \gamma^{ab} + {\cal K}(w) c^a c^b\, , 
\label{Tdef}
\ea
with
\be
{\cal K}^{-1}\equiv -2\frac{d{{\cal{L}}}}{ dw} >0.
\label{calKdef}
\ee
We define the norm of $c_a$ as
\be
\chi \equiv {c}^a {c}_a\, , 
\nn
\ee
so that the state parameter defined in Eq.~(\ref{wdef}) is given by
$ w =  {\cal K}^2  \chi$. 
One of the eigenvectors of $T^{ab}$ is $c^a$, whilst the other is 
$d^a \propto \epsilon^{ab} \varphi_{,b}$ (where $\epsilon^{ab}$ is the anti-symmetric tensor with $\epsilon_{01}=1$) with $c^a d_a =0$. Indeed, from Eq.~(\ref{Tdef}),
\ba
T^{ab} c_b &=& ({\cal L} + {\cal K}\chi) c^a,
\nn
\\
T^{ab} d_b &=& ({\cal L} ) d^a.
\nn
\ea
The corresponding eigenvalues correspond to the energy density $U$ or tension $T$ depending on whether $w$ is positive or negative:
\ba
&&U=-{\cal L}\, , \qquad \qquad \qquad T=-({\cal L}  + {\cal K} \chi) \qquad \qquad \text{for} \; w>0 \; ({\text{magnetic}})\, , 
\label{mess1}
\\
&&U=-({\cal L}  + {\cal K} \chi) \, , \qquad \; \; T=-{\cal L} \qquad \qquad  \qquad \; \; \; \; \; \text{for} \; w<0\; ({\text{electric}}).
\label{mess2}
\ea
Note that $U=U(\chi)$ and $T=T(\chi)$ so that $U=U(T)$, and hence elastic strings are characterised by a barotropic equation of state. (This can be determined explicitly in the case of the two Lagrangians given in Eqs.~(\ref{Le}) and (\ref{Lm}), see e.g.~\cite{Carter:1994hn}.) 
It follows from Eqs.~(\ref{mess1}) and (\ref{mess2}) that the relation
\be
U-T = {\cal K} |\chi|
\label{gen}
\ee
 holds whatever the sign of $w={\cal{K}}^2 \chi$, and that the tension of the string is always of lower value that $U$ (recall that ${\cal K}>0$).  As a result, the 
 transverse (or ``wiggle'') perturbations  on the string
propagate at speeds \cite{Carter:1989xk}
$c_{\rm E}^2 = {T}/{U}$ less than or equal to 1. There are also
 longitudinal (sound-like) perturbations, which travel at speed
$c_{\rm L}^2 = - {dT}/{dU}$.  For NG strings $c_{\rm E}^2=1=c_{\rm L}^2$, whereas for elastic strings, all field theory models studied in detail so far give $c_{\rm E}>c_{\rm L}$.
%\footnote{This restricts range $w$ for which the Lagrangians in (\ref{Le}) and (\ref{Lm}) are valid.}

It the following it will be more transparent to work in terms of four-dimensional `extrinsic' quantities, namely
\be
\overline{T}^{\mu \nu}= T^{ab} x^\mu_{,a} x^\nu_{,b} \, ,\qquad {\text{and }} \qquad \overline{c}^\mu = c^a x^\mu_{,a}.
\label{TCdef}
\ee
On using world-sheet reparametrisation invariance to choose the conformal gauge, $\dot x\cdot x'=0$, $x'^2=-\dot{x}^2$ (here $\dot{x}^\mu= \frac{\partial x^\mu}{\partial {\tau}}$,  $x'^\mu = \frac{\partial x^\mu}{ \partial \sigma}$) so that the induced metric $\gamma_{ab} =x'^2 {\rm{diag}}(-1,1)$,  it follows that the two vectors 
\be
u^\mu \equiv \frac{\dot{x}^\mu}{\sqrt{x'^2}}     \; , \qquad v^\mu \equiv \frac{{x'}^\mu}{\sqrt{x'^2}}
\label{uvdef}
\ee
with
\be
u^\mu u_\mu = -1 \, , \qquad v^\mu v_\mu = +1 \, , \qquad u^\mu v_\mu =0,
\nn
\ee
define a worldsheet orthogonal frame.  Furthermore, it is important to notice that provided $w\neq 0$ (thus excluding NG and chiral strings), one can use the freedom of Lorentz rotation on the worldsheet to choose $\varphi=\varphi(\tau)$ for electric strings, and $\varphi=\varphi(\sigma)$ for magnetic strings.  
In this {\it preferred rest-frame}, which we shall use repeatedly below, it is straightforward to show using Eqs.~(\ref{TCdef}) and (\ref{uvdef}), that the stress energy tensor is given by
\be
\overline{T}^{\mu \nu} = U u^\mu u^\nu  - T v^\mu v^\nu ,
\label{Tpreferred}
\ee
and that the components of the current $c^a$ take a particularly simple form, see (\ref{cdef}), from which $\overline{c}^\mu =\nu  u^\mu$ for electic strings, and $\overline{c}^\mu =\nu v^\mu$ for magnetic ones, where
\be
\nu = \frac{\chi}{\sqrt{|\chi|}} = ({\rm {sign}} (\chi))\sqrt{|\chi|}.
\nn
\ee

\section{Covariant Junction conditions for elastic strings}
\label{sec2}

We now study the dynamics of elastic strings meeting at a junction (itself assumed massless).
%~\footnote{\textcolor{blue}{The junction formation with a massive junction has been discussed in \cite{Avgoustidis:2014rqa}}}).  
We label the strings by an index $j$, so that for a V-junction $j$ runs from 1 to 2, and for a Y-junction $j$ runs from 1 to 3.
A V-junction is nothing other than a kink, or discontinuity in the tangent vector of the string. Since kinks are invariably 
formed when strings collide and create of Y-junctions, see \cite{Copeland:2006eh}, it is crucial to consider them in the following analysis.

Let string $j$ have coordinates $x^\mu_j(\sigma,\tau)$, world-sheet scalar field $\varphi_j$, Lagrangian density ${\cal L}(w_j)$, and corresponding tension $T_j$ and energy per unit length $U_j$. Furthermore the spatial world-sheet coordinate $\sigma$ is taken to increase towards the junction where it takes the value $s_j(\tau)$, thus
\be
-\infty < \sigma \leq s_j(\tau).
\nn
\ee
The action describing the dynamics of the system (strings and massless junction) is given by
\ba
S&=& \sum_{i}  \int  d  \tau  \, d \sigma \,\theta \left(s_{i}(\tau)-\sigma \right)  \sqrt{  -\gamma_i  } \;  {\cal L}(w_i)  
\nn
\\
&& +
 \sum_{i}  \int d \tau 
\left\{{\mathfrak{f}_i^\mu} \cdot [x_{i,\mu}(s_i (\tau),  \tau )-{X}_{\mu}(\tau)]+
{\mathfrak{g}_i} \cdot [\varphi_{i}(s_i (\tau),  \tau )-{\Phi}(\tau)] \right\} \; ,
\nonumber
\ea
where $\mathfrak{f}_{i}$ is  a 4-vector Lagrange multiplier imposing that the strings meet at the same position, namely at the junction
\ba
X^\mu(\tau) &=& x^\mu_i(s_i(\tau),\tau), 
\label{AtJ}
\ea
 and $\mathfrak{g}_i$ a scalar Lagrange multiplier imposing that the fields are continuous at the junction
 %\footnote{Other boundary conditions could be considered}, where they take the value $\Phi(\tau)$:
 \ba
 \Phi(\tau) &=& \varphi_i(s_i(\tau),\tau).
 \nonumber
\ea
Varying the action with respect to $X^\mu$  and $\Phi$ gives 
\be
\sum_i \mathfrak{f}^{\mu}_i=0=\sum_i \mathfrak{g}_i \; ,
\label{summ}
\ee
and with respect to $x^\mu_i$ and $\varphi_i$ gives
\ba
\partial_a \left(  \sqrt{  -\gamma_i  } \, T^{ab}_i x^\mu_{i,b} \theta \left(s_{i}(\tau)-\sigma \right)\right) &=& {\mathfrak{f}}_{i} ^{\mu}\delta( s_{i}- \sigma)\; ,
\nn
\\
\partial_a \left(  \sqrt{  -\gamma_i  } \, z^{a}_i \theta \left(s_{i}(\tau)-\sigma \right)\right) &=& {\mathfrak{g}}_{i} \delta( s_{i}- \sigma)\; ,
\nn
\ea
where $z^a_j = \sqrt{\kappa_{0,j}} c^a_j$.  Thus {\it at} the junction (where $\sigma = s_i(\tau)$), on using Eq.~(\ref{summ}), one deduces the conservation equations
\ba
\label{Xeom}
\sum_i \sqrt{  -\gamma_i  } \left(T^{0b}_i \dot{s}_i - T^{1b}_i \right) x^\mu_{i,b}  &=& 0,
\\
\label{phieom}
\sum_i \sqrt{  -\gamma_i  } \left(z^{0}_i \dot{s}_i - z^{1}_i \right) & = & 0.
\ea

In terms of 4-dimensional quantities defined in Eq.~(\ref{TCdef})
 these conservation equations can be written in a simpler and more physically obvious form \cite{Carter:1996mx}, namely
\bea
\sum_j\lambda_{j\nu} \overline{T}^{\nu \mu}_{j}&=&0,\label{eq:Tmunuconservation}\\
\sum_j\lambda_{j\nu} \overline{c}^{\nu}_j&=&0,
\label{eq:currentconservation}
\eea
where the $\lambda_j^\mu$, which we construct below, are outward directed unit normal vectors at the junction:
\be
\lambda^\mu_j  \lambda_{\mu,j}=+1, \qquad \lambda^\mu_j  \dot{{X}}_\mu=0.
\label{orthoJT}
\ee
Here the 4-velocity of the junction $\dot{{X}}^\mu(\tau)$ is obtained from Eq.~(\ref{AtJ}) as
\be
\dot{X}^\mu = \dot{x}^\mu_j + x'^\mu_j \dot{s_j} \, ,
\nonumber
\ee
where all quantities are evaluated at the junction and $ \dot{s_j} = {d s_j}/{d \tau}$.  The corresponding unit time-like vector $\dot{\hat{X}}^\mu(\tau)$ is thus
\be
\dot{\hat{X}}^\mu(\tau) = \frac{\dot{X}^\mu(\tau)}{N(\tau)} \, , \qquad \dot{\hat{X}}^2=-1, \qquad N^2(\tau)={\dot{X}^2}=x'^2_j(1-\dot{s}_j^2)
\label{Ndef}
\ee
which, in terms of the orthonormal unit frame vectors defined in Eq.~(\ref{uvdef}), becomes
\be
\dot {\hat{X}}^{\mu}=\Gamma_j\lb u_j^{\mu}+\dot s_j v^{\mu}_j\rb
\label{eq:junctionvelocity}
\ee
where $\Gamma_j \equiv (1-\dot{s}_j^2)^{-1/2}$.
It finally follows from Eq.~(\ref{orthoJT}) that
\be
\lambda_j^{\mu}=\Gamma_j\lb \dot s_j u_j^{\mu}+v^{\mu}_j\rb.
\label{eq:lambdaj}
\ee
Using this expression, together with the definitions of $\overline{T}^{\mu \nu}$ and $\overline{c}^\mu $ in (\ref{TCdef}) it is now straightforward to show the equivalence of `intrinsic' expressions for current and energy conservation at the junction Eqs.~(\ref{Xeom}), (\ref{phieom}), with their `extrinsic' form in Eqs.~(\ref{eq:Tmunuconservation}), (\ref{eq:currentconservation}).  

Before proceeding, we note that  Eqs.~(\ref{eq:junctionvelocity}) and (\ref{eq:lambdaj}) can be inverted to determine $u^\mu_j$ and $v^\mu_j$ at the junction, namely
\ba
u^\mu_j &=& \Gamma_j(\dot {\hat{X}}^{\mu} - \dot{s}_j \lambda^\mu_j)\, ,
\nn
\\
{v}^\mu_j &=&  \Gamma_j  ( \lambda^\mu_j - \dot {\hat{X}}^{\mu}\dot{s}_j).
\nn
 \ea
When the vectors $\dot {\hat{X}}^{\mu}$ and $\lambda^\mu_j$, as well as $\dot{s}_j$, are constants --- a particular case we will meet repeatedly below where we consider straight segments of string --- then it is straightforward to integrate these equations to determine the string position $x^\mu({\sigma,\tau})$. Indeed 
\bea
x^\mu_j({\sigma, \tau})& =&\Gamma_j \left[ \dot{\hat{X}}^\mu\,(\tau - \dot{s}_j \sigma) + \lambda^\mu_j\,(\sigma - \dot{s}_j \tau)\right]\, ,
\label{repassage}
\eea
where we have chosen the integration constant to be such that at the junction $x'^2_j=1$ (that is $N=1/\Gamma_j$) so that the induced metric is exactly Minkowski.

Now consider the junction condition Eq.~(\ref{eq:Tmunuconservation}). Working in the preferred rest-frame of each string where $ \overline{T}^{\mu \nu}_{j}$ takes the form given in Eq.~(\ref{Tpreferred}), it becomes 
\be
\sum_j \overline T_j^{\mu\nu}\lambda_{j\nu}=\sum_j \left[ 
%\Gamma_j^2\lb U_jv_j^2-T_j\rb\lambda_j^{\mu}-\Gamma_j^2v_j\lb U_j-T_j\rb\dot X^{\mu}
f_j\lambda_j^{\mu}+g_j \dot {\hat{X}}^{\mu}\right] = 0,
\label{eq:fjc}
\ee
where the 4D-Lorentz scalars $f_j$ and $g_j$ are given by
\ba
f_j &=& \Gamma_j^2\lb U_j\dot{s}_j^2-T_j\rb
\label{fdef}
\\
g_j &=& -\Gamma_j^2\dot{s}_j\lb U_j-T_j\rb .
%= -\Gamma_j^2\dot{s}_j {\cal K}_j |\chi_j|
\label{gdef}
\ea
Assuming that strings meeting at the junction are either all electric, $\overline{c}^\mu_j = \nu_j u^\mu_j$, or all magnetic $\overline{c}^\mu_j = \nu_j v^\mu_j$, the current conservation equation Eq.~(\ref{eq:currentconservation}) reduces, on using Eq.~(\ref{eq:lambdaj}), to
\bea
\sum_j \Gamma_j \nu_j \dot{s}_j &=& 0  \qquad {\text{(electric)}}\, ,
\label{Stare}
\\ 
\sum_j \Gamma_j \nu_j &=& 0 \qquad {\text{(magnetic)}}.
\label{Starm}
\eea

We now apply this formalism to study the dynamics of V and Y-shaped junctions, as well as to string collisions.

\section{V-shaped junctions}
\label{sec3}

For a V-shaped junction formed by two strings, $j=1,2$,  energy-momentum conservation at the junction (Eq.~(\ref{eq:fjc})) imposes\footnote{on contracting  Eq.~(\ref{eq:fjc}) with $\lambda^\mu_{1,2}$ and $\dot{\hat{X}}^\mu$}
\be
f_1= f_2=0,\qquad g_1=-g_2.
\nn
\ee
The first equality implies, from Eq.~(\ref{fdef}), that
\be
\dot{s}_j^2=\frac{T_j}{U_j}=c_{\mrm{E}{,j}}^2,
\label{SdotV}
\ee
so that junction propagates along the strings with extrinsic velocity $c_{\mrm{E}}$.  Hence, using Eq.~(\ref{gen}),
\be
\Gamma_j^{2}= \frac{U_j}{U_j-T_j} =\frac{U_j}{\cK_j |\chi_j|} \, ,
\nn
\ee
and condition $g_1=-g_2$ becomes (on using Eq.~(\ref{gdef})) 
\be
\sqrt{{U_1 T_1}}=\sqrt{U_2 T_2}.
%\label{eq:Vcondition1}
\nn
\ee
For NG strings ($U_i=T_i$) it follows that $U_1=U_2$, and the two strings meeting at the junction must be identical.
For a current-carrying string, the current conservation condition leads --- for both electric  (\ref{Stare}) and magnetic  (\ref{Starm}) strings --- to the same condition, namely
\be
\frac{{{\cal{L}}_1(w_1)}}{\cK_1(w_1)} = \frac{{{\cal{L}}_2}(w_2)}{\cK_2(w_2)}  \, ,
\nn
\ee
where ${\cK}(w)$ is defined in Eq.~(\ref{calKdef}).
Assuming the strings meeting at the kink are described by the same underlying field theory and hence the same underlying Lagrangian, current conservation therefore imposes that 
\be
w_1=w_2
\,
\nn
\ee
as expected --- namely the strings carry the same current.

For illustrative purposes, choose the first string to lie along the $x$-axis and the other to be the ($x,y$)-plane, so that in the V-junction rest frame (see figure 1)
\bea
\dot{\hat{X}}(\tau) &=& (1,0,0,0) \, ,
\nn
\\
\lambda^\mu_1 &=& (0,1,0,0) \, ,
\nn
\\
\lambda^\mu_2 &=& (0,-\cos\theta,\sin\theta,0).
\nn
\eea
\begin{figure}
\begin{center}
\includegraphics[width=12cm]{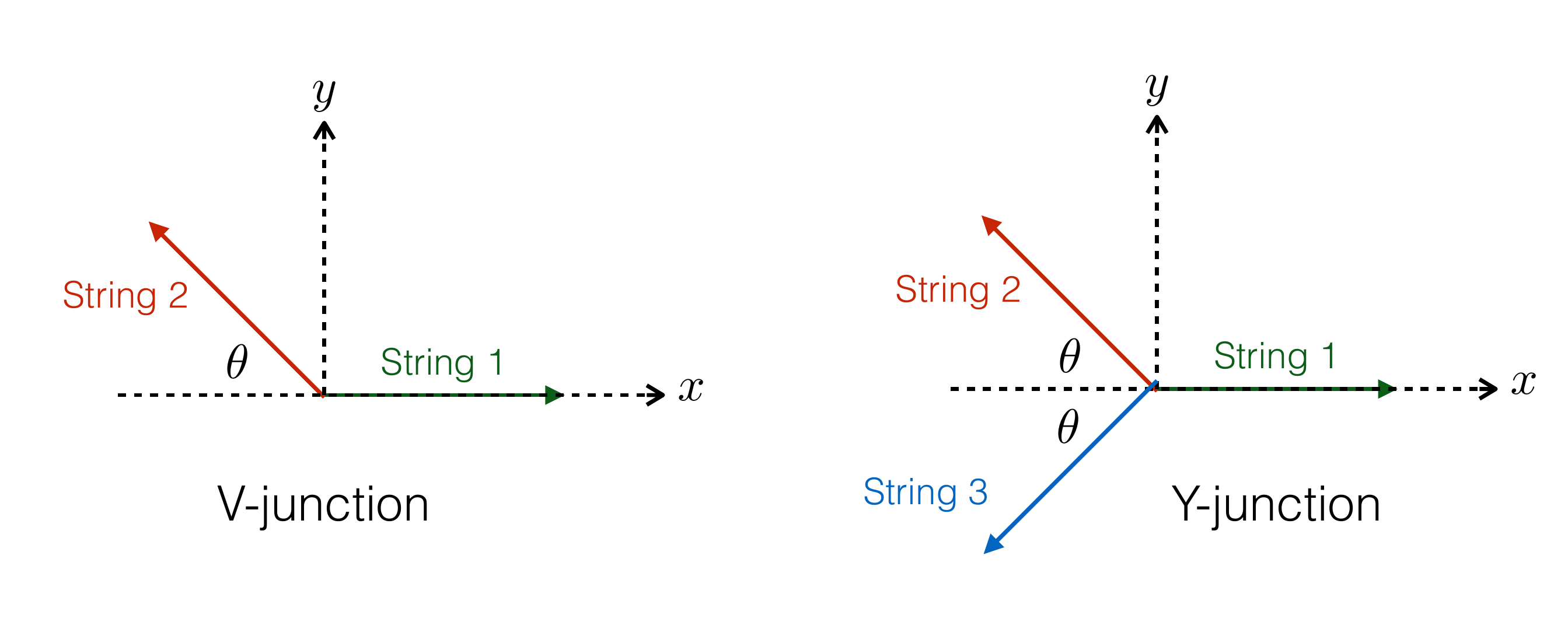}
\caption{Static V and Y junctions. The arrows indicate the direction of the outward pointing tangent vectors.}
\end{center}
\end{figure}
Since $w_1=w_2\equiv w$ are constants, $\dot{s_1}=-\dot{s}_2 =c_{\rm E}$ are also constant and given by Eq.~(\ref{SdotV}). Then from Eq.~(\ref{repassage}),
\bea
x_1^\mu(\sigma,\tau) &= &\Gamma_{\rm E}(\tau - c_{\rm E}\sigma,\sigma-c_{\rm E}\tau, 0 , 0).
\nn
\eea
Notice that the point $x_1^\mu(0,\tau)$ is not fixed but moves along string 1 in the negative $x$-direction with speed $c_E$. 
Physically, therefore, it is more transparent to work in a frame in which points of fixed $\sigma$ do not have a component of velocity along the string, namely in a frame in which $\dot{\vec{x}}\cdot\vec{x}'=0$. Hence, we now boost to a frame moving with velocity $-c_{\rm E}$ along the $x$-axis, and with a transverse velocity $v_z$ in in the $z$-direction: this will correspond precisely to the situation at hand when considering string collisions in section \ref{sec5}.
In this case
\bea
\dot{\hat{X}}^\mu&=& \Gamma_{\rm E}  (\gamma_z,c_{\rm E},0,\gamma_z v_z),
\label{sda}
\\
\lambda^\mu_1 &=& \Gamma_{\rm E} (\gamma_z c_{\rm E},1,0,\gamma_z v_z c_{\rm E}) \, ,
\nn
\eea
%\\
%\Longrightarrow \qquad 
so that from Eq.~\eqref{repassage},
\bea
x_1^\mu(\sigma,\tau) &=& (\gamma_z \tau,  \sigma, 0 , \gamma_z v_z \tau).
\label{theone}
\eea
When $v_z=0$ this reduces to $x_1^\mu(\sigma,\tau) = (\tau,  \sigma, 0 , 0)$ as required.

\section{Y-shaped junctions}
\label{sec4}

In this section we consider a Y junction where, as shown in Figure 1, for simplicity we assume that two of the strings deviate by the same angle $\theta$ from the direction of the first. Working in the junction rest frame, we thus have
\bea
\dot{\hat{X}}^\mu &=& (1,0,0,0)
\nn
\\
\lambda^\mu_1 &=& (0,1,0,0)
\nn
\\
\lambda^\mu_2 &=& (0,-\cos\theta,\sin\theta,0)
\nn
\\
\lambda^\mu_3 &=& (0,-\cos\theta,-\sin\theta,0)
%\label{big}
\nn
\eea
so that $\lambda_2^{\mu}+\lambda_3^{\mu}=-2\lambda_1^{\mu}\cos\theta$.  
The $y$ component  of the energy-momentum conservation Eq.~(\ref{eq:fjc}) gives
\bea
\Gamma^2_2(T_2 - U_2\dot{s}_2) &=& \Gamma_3^2(T_3 - U_3\dot{s}_3) 
\nn
\eea
where $\Gamma_i = \Gamma(\dot{s}_i)$.
This will be automatically satisfied if strings 2 and 3 are identical, which we will suppose in the following (and drop the subscripts 2 and 3). Then the $x$-component yields
\be
2 \cos\theta \,\Gamma^2 \lb T-U\dot{s}^2\rb = - \Gamma_1^2\lb T_1-U_1\dot{s}_1^2 \rb \, ,
\label{RI0}
\ee
which, in the case of NG strings, yields the condition on $\theta$ found in \cite{Copeland:2006eh,Copeland:2006if}, namely $\cos\theta=U_1/2U$ so that $U_1 \leq 2U$ as expected energetically.
Contracting  Eq.~(\ref{eq:fjc}) with $\dot{\hat{X}}^\mu$ gives
\be
2\,  \Gamma^2 \dot{s} \lb T-U \rb = - \Gamma_1^2\dot{s}_1 \lb T_1-U_1\rb ,
\label{RI1}
\ee
a condition which vanishes for NG strings. Finally, current conservation Eq.~(\ref{Stare}) gives
\bea
2\, \Gamma\, \dot{s}\, \nu &=& -\Gamma_1 \dot{s}_1 \nu_1\qquad (\text{electric}),
\label{RI2}\\
2\, \Gamma\,\nu &=& -\Gamma_1\,\nu_1\qquad (\text{magnetic}).
\label{RI2m}
\eea
Let us assume that the Lagrangian of each string is known, see for instance Eq.~(\ref{Le}), as well as the current on that string, namely the values of $(w,w_1)$.  (That is, the tension and energy density of each string is known.) Then the unknown variables describing the junction are $(\dot{s},\dot{s}_1,\cos\theta)$. The first two are determined from Eqs.~(\ref{RI1}) and (\ref{RI2}), and $\cos\theta$ from Eq.~(\ref{RI0}).  For instance, for electric strings  it follows from Eqs.~(\ref{RI1}) and (\ref{RI2}) that
%\textcolor{blue}{
\bea
\dot{s} &=& - \frac{\cK(w)}{2\cK(w_1)}\dot s_1 \, , \qquad  \dot{s}^2 = \frac{|w_1| - |w|}{|w_1| -|w| \frac{4 \cK^2(w_1)}{\cK^2(w)}}\qquad (\text{electric}),
%\dot{s} &=& - \frac{2\cK(w_1)}{\cK(w)}\dot s_1 \, , \qquad  \dot{s}^2 = \frac{|w_1| -|w| \frac{4 \cK^2(w_1)}{\cK^2(w)}}{|w_1| - |w|}\qquad (\text{magnetic}),
\nn
\eea
which implies that $|w_1|>|w|$ and $2 \cK(w_1) \leq \cK(w)$ where, for the explicit electric string model of (\ref{Le}), $\cK (w)= 1-{|w |}/{M^2}$.
The condition $\cos\theta\leq1$ further restricts the parameter space.%, as shown in figure ?.

Finally, the position of each string can be obtained from (\ref{repassage}) as before.

\section{String collisions}
\label{sec5}

We now consider a string collision between two incoming and identical strings at angles $\pm \alpha$ in the $(x,y)$ plane with velocities $\pm v_z$ in the $z$-direction, as shown in figure 2.  The values of $\alpha$ and $v_z$, as well as the type of incoming string, namely its tension $T$ from which $U(T)$ is determined (or equivalently ${\cal L}(w)$ and $w$) are thus given initial conditions for this problem.  Note that this means that $c_{\rm E}$, the speed of any V-junction formed, is also therefore known from the initial conditions.
\begin{figure}
\begin{center}
\includegraphics[width=12cm]{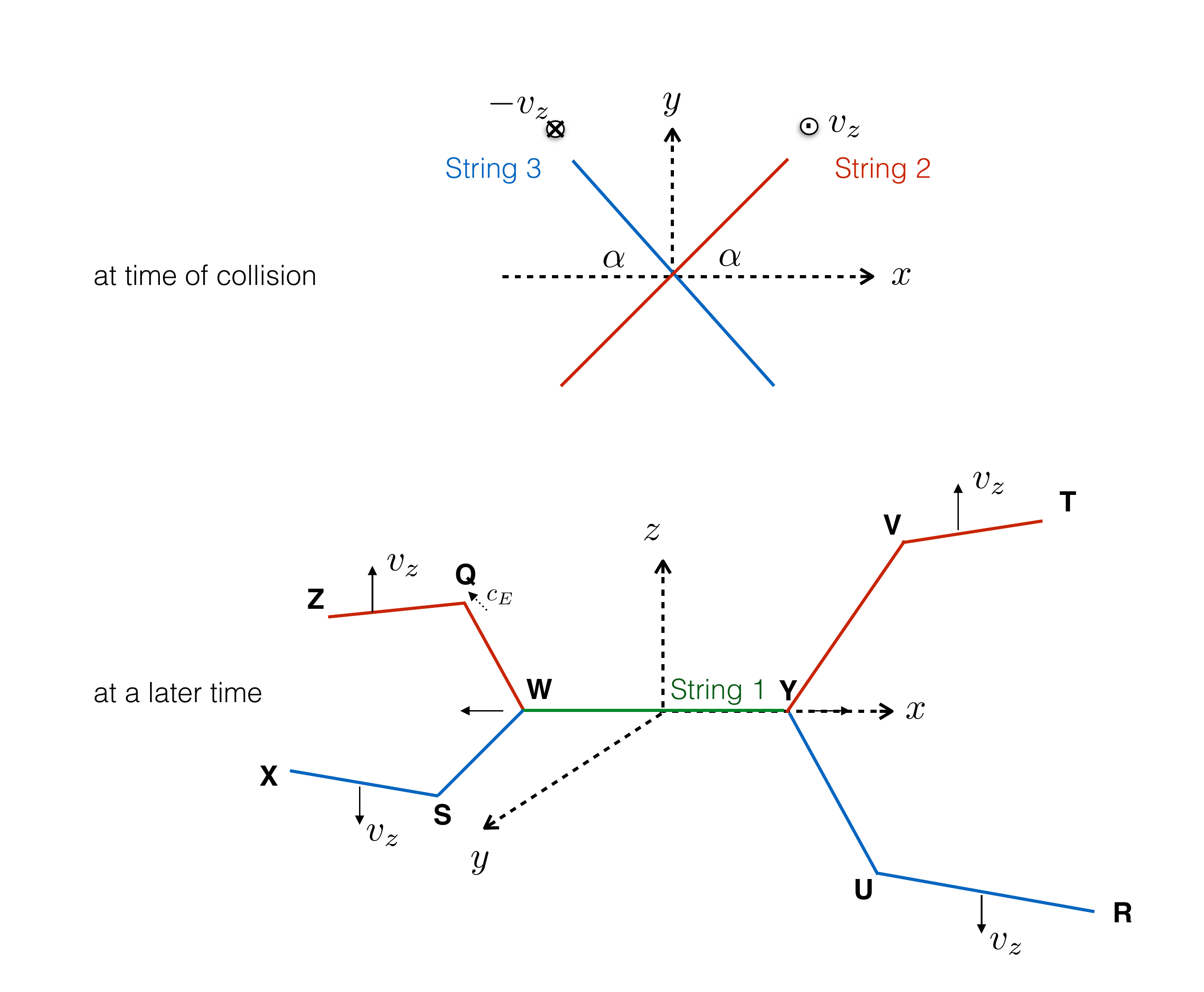}
\caption{String collision.  Before the collision two identical are parallel to the $(x,\,y)$ plane forming an angle $\pm \alpha$ with the $x$-axis and moving in the $z$-direction with velocities $\pm v_z$.  After the collision, two Y-junctions (labelled $W$ and $Y$) moves along the $x$-axis while four V-junctions (labelled $Q,S,U,V$) travel along the strings at velocity $c_{\mrm E}$ in the string's preferred frame. }
\end{center}
\end{figure}

There are at least 3 different possible outcomes of such a collision: i) the strings could cross without interaction, ii) they could intercommute, or iii) the two initial strings could become joined by the formation of connecting segment (string 1 in the figure) which, by symmetry, is at rest in the $x$ direction.  It is scenario iii) that is envisaged here.

For NG strings (when $U=T$), it is has been show in \cite{Copeland:2006eh,Copeland:2006if} that the formation of a connecting segment is only possible in a well defined region of the $(\alpha,v_z)$ plane. This kinematic constraint comes from the physical requirement that the length of the connecting segment should increase in time. Our aim in this section is to see if a similar constraint exists for current carrying string collisions.

After the collision, two Y-junctions (labelled W and Y in figure 2) move along string 1 parallel to the $x$-axis, while four V-junctions (labelled Q,S,U,V) travel along the strings with velocity $c_{\mrm E}$. By causality, the segments ZQ and VT of the string 2 remain unperturbed (and  analogously for string 3), and are given by an expression similar to that of Eq.~(\ref{theone}), namely
\be
x_{2}^{\mu}(\sigma,t)=\gamma_z\lb \emu_t+ v_z\,\emu_z\rb t+\lb  \cos\alpha\,\emu_x+\sin\alpha\,\emu_y\rb\sigma,
\nn
\ee
where the unit vectors $\emu_t=(1,0,0,0)$, $\emu_x=(0,1,0,0)$ and analogously for $\emu_y$ and $\emu_z$.
The unit-velocity 4-vector of the $V$-junction V is then given by, see Eq.~(\ref{sda}),
\be
\dot{\hat{X}}^\mu_V(t) =
 \Gamma_{\rm E} \gamma_z \left(\emu_t +v_z\emu_z \right) + \Gamma_{\rm E} c_{\rm E}(\cos\alpha \emu_x + \sin\alpha\emu_y) .
 \nn
\ee
The corresponding vector $\dot{\hat{X}}^\mu_U(t)$ for the junction U is simply obained from the above by flipping $v_z \rightarrow -v_z$ and $\alpha \rightarrow -\alpha$.

On the other hand, string 1 is given by
\be
x_{1}^{\mu}(\sigma,t)=t \emu_t +\sigma \emu_x,
\nn
\ee
and at the $Y$-junction Y, $\sigma = {s}_1(t)$. Hence
\bea
\dot{\hat{X}}_Y^\mu(t) &=&
 \Gamma_1  \left( \emu_t  +  \dot{s}_1\emu_x \right),
 \nn
 \\
 \lambda^\mu_1 &=& \Gamma_1  \left(\dot{s}_1  \emu_t  +  \emu_x \right).
 \nn
\eea
Junctions V and Y therefore have a relative speed $v_+$ with $\gamma_+ = (1-v_+^2)^{-1/2}$ given by
\be
\gamma_+ = - \dot{\hat{X}}_Y^\mu (t)\dot{\hat{X}}_\mu^V(t)  = \Gamma_1  \Gamma_{\rm E}\left(\gamma_z - \dot{s}_1 c_{\rm E} \cos\alpha\right).
\label{gamplus}
\ee
In order to apply the junction conditions at the $Y$-junction Y, we need to construct outward pointing tangent vectors on strings 2 and 3 at Y.
On string 2, $\lambda_2^\mu$ is a linear combination of $\dot{\hat{X}}_Y^\mu$ and $\dot{\hat{X}}^\mu_V$, with coefficients determined by the normalisation of $\lambda_2^\mu$ and its orthogonality with $\dot{\hat{X}}_Y^\mu$: 
\bea
\lambda_2^\mu &=& -(v_+ \gamma_+)^{-1}\left(  \dot{\hat{X}}^\mu_V - \gamma_+ \dot{\hat{X}}_Y^\mu \right).
\label{lam2}
\eea
The expression for $\lambda_3^\mu$ is the same, but with $ \dot{\hat{X}}^\mu_V$ replaced by $\dot{\hat{X}}^\mu_U$.
%\bea
%\lambda_3^\mu &=& (v_+ \gamma_+)^{-1}\left(  \dot{\hat{X}}^\mu_U - \gamma_+ \dot{\hat{X}}_Y^\mu \right)
%\eea
Finally (up to an irrelevant overall additive constant) from Eq.~(\ref{repassage}), the string on segment YV is given by\footnote{Assuming that $\dot{\hat X}_Y^\mu$, $\lambda_2^\mu$, and $\dot s_2$ are constants, as expected for straight strings}  
\be
x_{{\rm YV}}^\mu(\sigma,t) = \Gamma_2\left[\dot{\hat{X}}_Y^\mu (t - \dot{s}_2\sigma) + \lambda_2^\mu (\sigma - \dot{s}_2t) \right] \, ,
\label{x22}
\ee
where $\sigma=\dot{s}_{2}t$ at the Y-junction.  This string had better connect the V and Y-junctions!  Or, in other words, the value of $\dot{s}_2$ must such that at the V-junction where $\sigma=c_{\rm E} t$, the equality $\frac{d}{dt}(x_{\rm{YV}}^\mu(c_{\rm E} t,t)) =  \dot{{X}}^\mu_V$ holds.  Hence, since $\dot{\hat{X}}^\mu_V = \Gamma_{\rm E} \dot{{X}}^\mu_V$, it follows from Eq.~(\ref{x22}) that
\be
\dot{\hat{X}}^\mu_V =\Gamma_{\rm E} \Gamma_2 \left[ \dot{\hat{X}}_Y^\mu (1-c_{\rm E}\dot{s}_2) + \lambda_2^\mu (c_{\rm{E}} - \dot{s}_2)  \right].
\label{x23}
\ee
Combining Eqs.~(\ref{lam2}) and (\ref{x23}) therefore gives
\be
\dot{s}_2 = \frac{ c_{\rm E}+v_+}{1+c_{\rm E} v_+} = \dot{s}_3 \equiv \dot{s},
\label{sdotT}
\ee
where $v_+=-\left({1-\gamma_+^{-2}}\right)^{1/2}$.
Finally, the angle $\theta$ between strings 1 and 2 is given by
\be
\cos\theta = -\lambda^\mu_1  \lambda_{2,\mu} = \frac{\Gamma_1 \Gamma_{\rm E}}{v_+ \gamma_+} \left(\dot{s}_1 \gamma_z - c_{\rm E} \cos\alpha \right)
= \frac{1}{v_+} \frac{(\dot{s}_1 \gamma_z - c_{\rm E} \cos\alpha)}{(\gamma_z - \dot{s}_1 c_{\rm E} \cos\alpha)}
\label{costheta}
\ee
where $v_+$ itself obtained from Eq.~(\ref{gamplus}).

We are now finally in a position to apply the $Y$-junction conditions Eqs.~(\ref{RI0}) to (\ref{RI2}) to Y.    First, however, let us understand the NG limit. Here there is obviously no current conservation condition, and furthermore one of the other stress-energy conservation equations (namely Eq.~(\ref{RI1})) is trivially satisfied since $U=T$ for NG strings.  How does the resulting condition reduce to the kinematic constraint of \cite{Copeland:2006eh}?  The NG limit is obtained when $c_{\rm E} \rightarrow 1$, and from Eq.~(\ref{gamplus}) that obviously implies $v_+\rightarrow -1$ as expected. Then from Eq.~(\ref{costheta}),
\be
\cos\theta \rightarrow \frac{\dot{s}_1 \gamma_z - \cos\alpha}{\gamma_z - \dot{s}_1 \cos\alpha} \qquad {(\text{NG limit})}.
\nonumber
\ee
The only remaining junction condition is Eq.~(\ref{RI0}), which for NG strings reads
\be
2\cos\theta\, T = -T_1.
\nonumber
\ee
On substituting for $\cos\theta$ one arrives directly at the result
\be
\dot{s}_1 = \frac{ 2T\gamma_z^{-1}\cos\alpha- T_1 }{2T - T_1\gamma_z^{-1}\cos\alpha }.
\nonumber
\ee
Since the length of the joining string must increase, $\cos\alpha > \gamma_z T_1/(2T)$.  Hence for a given $T_1$ and $T$, the formation of Y-junctions in such a collision can only occur in a certain region of the $(\alpha,v_z)$ plane (namely small $\alpha$ and $v_z$), as discussed in depth in \cite{Copeland:2006eh}.

Now let us turn to current carrying string collisions, for which the three junction conditions (\ref{RI0}) to (\ref{RI2}) must be satisfied. Relative to the static Y-junction considered in section \ref{sec4}, 
 it is important to notice two crucial differences.  The first is that $\dot{s}$ is not free variable: from Eq.~(\ref{sdotT}), given $\alpha$, $v_z$ and $c_{\rm E}$ (known from the initial conditions), it is a function of $\dot{s}_1$.  Secondly, the angle $\cos\theta$ is also no longer a free variable but determined by $\dot{s}_1$.  Therefore, contrary to the case considered in section \ref{sec4} where there were 3 unknown variables,  there are only {\it two} unknown variables in this problem involving colliding strings namely $(\dot{s}_1,w_1)$. However, there are 3 junction conditions!   The system is therefore over-determined. Indeed, we arrive at the conclusion that in such a collision, if a joining string forms, then it can {\it not} be described by an elastic string model (that is, by a barotropic equation of state). We expect that there must be time dependence and dissipative processes at the Y-junctions, which will generically require the use of a non-conservative model.  It is also of great interest to investigate numerically, as will be reported elsewhere \cite{Hiramatsu}.

\section{Conclusions}
\label{sec6}

In this paper we have extended the analysis of string collisions, with the subsequent formation of Y-junctions, to
{\it elastic} string models which characterise current-carrying strings.  To do so, we have developed a fully covariant formalism. We have shown that when Y-junctions form, the resulting system of equations is overdetermined: the number of the unknown variables is smaller than the number 
of junction conditions.  
Therefore our main 
result is that the collision of current-carrying strings in which a joining string forms cannot be described by the elastic string models,
and a non-conservative model must be used. This will be the subject of future work, together with a numerical study of
the collision of field theory current-carrying strings in the $U(1)\times U(1)$ model.

\acknowledgments

We are particularly grateful to Brandon Carter who played a very important r\^ole in initiating this work many years ago with D.A.S., and whose encouragement enabled us to finish it.
D.A.S thanks CERN and the Fondation Eva Maier for hospitality whilst this work was being finished, and Patrick Peter for useful discussions.
This work was supported in part by Grant-in-Aid from
the Ministry of Education, Culture, Sports, Science and Technology (MEXT) of Japan,  Nos.~17K14304 (D.Y). T.~H. is supported by JSPS Grant-in-Aid for Young Scientists (B) No.~16K17695.

\bibstyle{aps}
\bibliography{brandon}

\end{document}